\documentclass[12pt]{article}
\begin{document}
\begin{center} \bf \large
Possibility of Sound Propagation in Vacuums with the Speed of Light
\\ \vspace{2cm}\normalsize 
Robert Lauter \\
Max-Planck Institut f\"ur Kolloid- und Grenzfl\"achenforschung, 
Max-Planck Campus, Haus 2, Am M\"uhlenberg 2, 14476 Golm (Potsdam)
\end{center}

An important question of theoretical physics is whether sound is able to propagate in vacuums
 at all and if this is the case, then it must lead to the reinterpretation of one zero restmass 
particle which corresponds to vacuum-sound waves. 
Taking 
the electron-neutrino as the corresponding particle, its observed non-vanishing rest-energy may
 only appear for neutrino-propagation inside material media. The idea may also influence the physics
 of dense matter, restricting the maximum speed of sound, both in vacuums and in matter, to the
 speed of light.
\vspace{1cm} \\
PACS numbers: 03.30.+p; 14.60.Pq; 43.35.Gk; 97.60.Jd
\vspace{1cm} \\
Introduction:
\vspace{.5cm} \\
Since the idea of sound propagation in vacuums is generally rejected, I will begin by discussing the various reasons for which this idea is not believed by the great majority of physicists.
The most obvious reason is that the transmission of normal acoustic sound waves is, in contrast to electromagnetic waves, not observed in vacuums. Nevertheless, this is not an argument entirely against sound propagation in vacuums, since sound waves correspond to zero restmass particles which have to propagate with velocity c in vacuums. Therefore, from the very high ratio of the phase velocity of sound in vacuums and, for example, in air of approximately $10^6$, 
one has to expect negligible transmission of normal acoustic sound waves into vacuums even when sound propagation in vacuums exists.
 
The second reason why it is not believed that sound propagates in vacuums, is that compressionable waves do not exist in vacuums. However, the criterion compressibility does not determine which thermodynamical parameter of state is altered during a particular propagation of sonic waves. Hence, there are some examples of acoustic wave propagation which occur without altering the pressure, the density, or the volume of the medium. The most ordinary case is the propagation of transversal sound waves which occur without altering the volume of the medium. The second example is the propagation of the so called 
``second sound'' and the third one, the so called ''zero sound''. The two latter types of acoustic phenomena only occur inside superfluid matter, the first of which represents periodic oscillations of the temperature without altering the pressure (while the density is altered) while zero sound  propagates without altering the density of the medium (while pressure changes). Thus, in material media, it depends on the actual physical circumstances inside the medium as to which parameters of state are changed during sound propagation, and neither changes of pressure or volume, nor changes in density are necessary preconditions for sound propagation in general. It is therefore doubtful if compressibility is a necessary property of a medium to permit sound propagation in general.

The third argument against sound propagation in vacuums is the same as that used during the discussion about the ether-hypothesis in the last century. Most people, including physicists, could not imagine any wave propagation without the displacement of particular components (atoms, molecules) inside a medium. But since there is no doubt that electromagnetic waves propagate in vacuums without the oscillation of any material medium through which they are propagating, the same question also has to be asked for acoustic waves.

In the following investigation I shall first derive some basic properties of vacuum-sound waves from the thermodynamical properties of a vacuum. This is possible because thermodynamics do not make any restriction concerning the internal structure of matter. Therefore the whole thermodynamics remained true even if matter was a continuum (which we will treat the vacuum as). Another approach is the investigation of vacuum-sound waves by means of lattice-theory. 
Secondly the results of this investigation are compared to the electron-neutrino's properties, which seem to fit the properties of vacuum-sound waves.
\vspace{.5cm} \\
Investigation of the vacuum's properties, concerning sound propagation:
\vspace{.5cm} \\
The thermodynamical investigation of sound-propagation in vacuums requires at first a classical, i.e. non-quantized, definition of a vacuum. I shall adopt the definition of T. H. Boyer [1] who states that a vacuum is a region of space from which all rest-energy and additionally all thermal radiation has been removed. This region of empty space still contains the zero-point energy which comes from the fluctuating fields of the strong and weak nuclear as well as the electromagnetic force, which can not be removed 
[2]. \linebreak[4]
Thus, propagation of sound in vacuums is considered as occurring at zero K like zero sound in material media. 
The above classical definition of a vacuum consequently leads to vanishing values for the pressure and the total energy density of a vacuum, or $p \rightarrow 0$ and 
$\epsilon \rightarrow 0$. 
The relativistic relationship due to the velocity of sound propagation according to Bludman and Ruderman [3] is:

\begin{equation} c^2_s = c^2 \cdot (\frac{dp}{d \epsilon})_s
\end{equation}
with: \\
$c_s$ speed of sound \\
$c$ speed of light \\
$p$ pressure \\
$\epsilon$ total energy density (including rest energy) \\
$S$ entropy.
\vspace{1cm} \\
This equation was derived by replacing the density, $\rho$, 
by the energy-density. This is possible because of the mass-energy equivalence which resulted from the theory of special relativity. It shows that rest-energy and hence matter, as a supporting medium, is not a necessary precondition for propagation of sonic waves in general, hence sound propagation in vacuums is permitted due to the theory of special relativity. It also shows that $c_S$ equals $c$ 
if, as in the case of a vacuum, the following equation is fulfilled:

\begin{equation} 
\lim_{dp \rightarrow 0, d\epsilon  \rightarrow 0} (\frac{\partial p}{\partial \epsilon})_{s} =1
\end{equation}

According to Bludman and Ruderman [3], Lorentz-invariance imposes no restriction on the speed of sound in material media (this can also be explained by the inapplicability of the Lorentz-transformation to zero-restmass particles, since they have to propagate with velocity c in a vacuum). Thus, the principle of sound propagation in vacuums provides a limiting condition for the velocity of signal transmission. Bludman's and Ruderman's treatment proved that a material system which became ultrabaric, i.e. p exceeds 
$\epsilon$, had to become superluminal before, i.e. dp exceeds $d\epsilon$. 
This is only possible in the case of one-dimensional pressure-fluctuations, since otherwise $p \leq \frac{1}{3}\epsilon$ (the equality holds for a photon-gas). Thus I want to investigate the case of a one-dimensional photon-gas in order to show that a vacuum which can be constructed  according  to the above  given  definition  will  yield $p \rightarrow 0$ and $\epsilon \rightarrow 0$  while $p = \epsilon$ always holds true.
 
Consider a system of non interacting photons which are moving in parallel direction between two ideally reflecting surfaces perpendicular to the direction of the movement of the photons. The pressure on the two plates can be calculated from the momentum which is applied per time-interval on the two plates. For one photon this yields:

\begin{equation} 
p = \frac{F}{A}
\end{equation}
$p$ pressure \\
$F$ Force\\
$A$ surface-area\\

and:

\begin{equation} 
F= \frac{dP}{dt}
\end{equation}
$P$ momentum\\
$t$ time\\

Since we considered non-interacting photons and ideal reflection we can also write:

\begin{equation} 
F = \frac{\Delta P}{ \Delta t}
\end{equation}

The time which is needed by a photon to return to the same position after two reflections is:
\begin{equation} 
\Delta t= \frac{2l}{c}
\end{equation}
$l$, distance between the two plates
\vspace{1cm} \\
Consequently: 
\begin{equation} 
p= \frac{c \Delta P}{2lA}
\end{equation}

At each reflection a photon transfers twice its momentum to a plate, thus:
\begin{equation} 
p= \frac{cP}{lA}=\frac{E}{V}= \epsilon
\end{equation}
$E$, energy of one particle\\
$V$, Volume\\

For a system of $N$ particles this gives: 
\begin{equation} 
p= \epsilon= \frac{\sum_{N=1}^{N=N} E_N}{V}
\end{equation}
$N$, number of particles
\vspace{1cm} \\
In this regime the vacuum can be constructed by two approaches. Either the number of particles approaches zero or the volume approaches infinity. In both cases $p = \epsilon$ always holds true. 
Thus  it  is possible  to apply  thermodynamical  functions to  a  vacuum.  Since  
$ p \rightarrow 0$   
and $dp \rightarrow 0$ in vacuum, sonic excitations must represent isobaric changes of state, or 
$dp = 0$. The heat which is changed in an isobaric process is the enthalpy, H. 
\vspace{1cm} \\
The total differential of the enthalpy is:
\begin{equation}
dH = dU + pdV + Vdp
\end{equation}
$U$, internal energy
For isobaric changes of state this reduces to:
\begin{equation} 
dH = dU + pdV
\end{equation}

Because of  its Lorentz-invariance, every zero-restmass field which propagates in a vacuum must represent  isoentropic (or adiabatic)  changes  of  state.  Since 
$ p \rightarrow 0$,  no work is 
applied to the vacuum for sonic excitations and since for adiabatic changes of state:

\begin{equation} 
\delta W = \delta Q
\end{equation}
$W$, work \\
$Q$, heat
\vspace{1cm} \\
It follows that:
\begin{equation} 
dH = dU = 0
\end{equation}

\begin{equation} 
dU = (\frac{\partial U}{\partial T})_V dT +(\frac{\partial U}{\partial V})_T dV = C_V dT + pdV
\end{equation}

and , since $p = 0$

\begin{equation}
dU = (\frac{\partial U}{\partial T})_p dT = C_V dT
\end{equation}
\vspace{1cm} \\
which reproduces the condition that the temperature of a vacuum  always remains zero K.
(This means that empty space can be heated up yielding a thermal, in addition to its zero-point spectrum, hence a vacuum is equivalent to empty space at zero K. 
Briefly summarising the results obtained until now, one can conclude that no thermodynamical parameter of state is changed during sound-propagation in vacuums. This result is obtained because we used a classical description of a vacuum. 
Considering quantum-theory, the particle corresponding to sound waves carries a momentum as well as energy. The energy is added to a vacuum merely for time-intervals which are permitted due to the uncertainty-principle while the momentum of zero restmass particles is a global property of the corresponding wave-field since individual particles are not localizable. But since the corresponding wave-field carries a momentum, a wave of determined internal symmetry (longitudinal or transversal) would also apply a pressure on the vacuum because in this case maxima and minima of pressure would have to appear. Thus we additionally have to assume that the vacuum's sonic wave field must be a scalar one.

The theory of wave-propagation in periodic structures (lattices) shows another property of the vacuum. This theory shows that waves only suffer from dispersion (i.e. the phase-velocity is a function of the energy) when the wave length is comparable to twice the distance between lattice-points [4]. No dispersion is observed for wavelengths, which are large compared with the distance between lattice-points. Thus, no dispersion for all wavelengths only agrees with vanishing distances between lattice-points which is generally known as the continuum-limit. This implies that the vacuum's sonic waves suffer from dispersion when they propagate in material media and hence they obtain a virtual restmass, like photons which propagate inside optical dense media. 
Since I want to compare the above derived properties of vacuum-sound waves to the properties of the electron-neutrino, I want to resume the obtained results.
We have to expect the following properties of vacuum-sound waves:

\begin{itemize}

\item Sound waves 
correspond to zero-restmass particles which have to propagate with velocity c in a vacuum.
 \item
The corresponding zero-restmass-particle should be emitted only from matter whose density is high enough to permit sound-propagation near velocity c inside the material. This restricts emission of vacuum-sound waves to material of nuclear density. The high energy which is liberated in nuclear processes causes an equivalent short wavelength of the emitted particles. 
\item
Thus, these sound waves should interact negligibly with matter of normal density, e.g. air, due to the high ratio of the phase-velocity of sound propagation in a vacuum and in air of approximately 
$10^6$ and because of their very short wavelength which is smaller than double the lattice-spacings 
of normal materials.

\item 
Propagation of those sound waves inside material of suitable phase-velocity should lead to a non-vanishing virtual restmass as in the equivalent case of photons when they propagate through optical dense media. 
\item
The corresponding elementary particle should correspond to a scalar wave-field.
\item The corresponding particle should carry no electric charge.
\end{itemize}

As far as it is known to the author, the electron-neutrino 
$(\nu_{e}, \bar{\nu}_{e})$ 
represents the unique zero-restmass particle which corresponds to a scalar wave-field and is additionally electrically neutral. Thus we now want to consider this elementary particle as corresponding to the vacuum-sound wave's zero-restmass particle. Other theoretical evidence which agrees with this assumption represents the description of neutrino-emission in neutron-stars by statistical physics. Ruderman et al. [5] showed that below the transition-temperature of the neutron fluid to the superfluid state, pairs of excited neutron quasiparticles may recombine, resulting in the emission of neutrino-antineutrino pairs. This process is similar to the phonon emission by recombination of quasiparticles in conventional superconductors and hence any difference between the emission of sound-quanta (phonons) and the emission of neutrino-antineutrino pairs cannot in this case be distinguished. 

Furthermore, neutrinos are emitted only from matter of nuclear or even higher density in which the speed of sound approaches the speed of light. They also have very short wavelengths due to their generation in high energetic, nuclear processes (neutrinos are, in a sense, the $\gamma$-quanta of 
sound) which causes them to be transmitted through materials of normal density.

Actually, there is some discussion as to whether the observed rest-energy of the electron-neutrino appears only for neutrino-propagation inside matter or also in a vacuum. A non-vanishing rest-energy of this particle would enable it to be transformed into different kinds of neutrinos,  
$\nu_\mu$ and $\nu_\tau$ [6]. Until now it is not clear whether this transformation is possible also in a vacuum or only when the particle propagates in material media [6]. The latter case is similar to that of photons, since photons also omit no rest-energy when they propagate in a vacuum but obtain a rest-energy when they propagate in optical dense media. This leads to the proposal of an experiment which is, in principle, able to show, whether neutrinos propagate in a vacuum as zero-restmass particles or as non zero-restmass particles:
If neutrinos are zero-restmass particles when they propagate in a vacuum, a detector which approaches the sun will recognise an intensity-increase which is exactly proportional to $d^{-2}$ 
($d$ = distance of the detector from the sun). 
\vspace{1cm} \\
Conclusion:
\vspace{1cm} \\
Firstly, the possibility of sound-propagation in vacuums implies an important consequence for the cooling of neutron-stars by neutrino-emission. If neutrinos are the corresponding zero-restmass particles to vacuum-sound waves, a star whose matter remains in a state which approaches the velocity c for sound propagation is able to emit internal energy of heat by the emission of its acoustic excitations in form of neutrinos. This cooling-mechanism has to be considered as an important process for the calculation of the stellar collapse of stars which have exhausted their nuclear fuel. 

Secondly, the idea of sound propagation in vacuums distinguishes clearly between Einstein's and Lorentz's classical interpretation of the principle of relativity.  Lorentz's interpretation that there exists only one distinguished inertial frame, that one in which the speed of light is isotropical c, permits the existence of an ``ether'' [7] (the medium's properties of empty space) which according to J. Brandes [7] is every physical property which empty space has additionally to its volume.
Generally it is assumed that there is no measurable consequence between both theories of gravitation. But since Lorentz's attitude allows for the concept of an "ether", this concept is really a necessary consequence of his theory. Thus, the necessary existence of an ether and hence of sound waves which propagate inside this medium of empty space, consequently requires the existence of a zero restmass particle which corresponds to these sonic waves. For this particle, only zero rest-energy is permitted when it propagates through a vacuum. If neutrinos can be interpreted as the corresponding particle, the only measurable consequence which leads to a decision between Einstein's and Lorentz's attitude may be its rest-energy in vacuums.
\vspace{1cm} \\
Acknowledgement: I wish to thank all my colleagues who contributed to this paper 
through engaged discussion.
\vspace{1cm} \\
References: 
\vspace{1cm} \newline
[1]    T. H. Boyer, Scientific American, 70, (August 1968). \newline
[2]    P. Yam, Scientific American, 54, (December 1997).\newline
[3]    S. A. Bludman, M. A. Ruderman, Phys. Rev. 170, 1176 (1968).\newline
[4]    L. Brillouin, Wave Propagation in Periodic Structures, (Dover Publica-

tions, 1953) p. 4.\newline
[5]    M. Ruderman, E. Flowers, P. Sutherland, Ap.J. 205, 541 (1976).\newline
[6]    G. Drexlin, Phys. Bl. 55/2, 25 (1999/2).\newline
[7]    J. Brandes, Die relativistischen Paradoxien und Thesen zu Raum und 

Zeit: Interpretationen  der  speziellen  und  allgemeinen  Relativit\"atstheorie,  

(Verlag  relativistischer Interpretationen, 1995) p. 176 ff.
 
\end{document}